# Edge-functionalization of armchair graphene nanoribbons with pentagonal-hexagonal edge structures

Junga Ryou[1,2], Jinwoo Park[1], Gunn Kim[1] and Suklyun Hong[1]

[1]Department of Physics and Graphene Research Institute, Sejong University, Seoul 05006, Republic of Korea

[2]Korea Research Institute of Standards and Science, Daejeon 34113, Republic of Korea

Using density functional theory calculations, we have studied the edge-functionalization of armchair graphene nanoribbons (AGNRs) with pentagonal-hexagonal edge structures. While the AGNRs with pentagonal-hexagonal edge structures (labeled (5,6)-AGNRs) are metallic, the edge-functionalized (5,6)-AGNRs with substitutional atoms opens a band gap. We find that the band structures of edge-functionalized (5,6)-N-AGNRs by substitution resemble those of defect-free (N-1)-AGNR at the Γ point, whereas those at the X point show the original ones of the defect-free N-AGNR. The overall electronic structures of edge-functionalized (5,6)-AGNRs depend on the number of electrons, supplied by substitutional atoms, at the edges of functionalized (5,6)-AGNRs.

# 1. Introduction

Since its first isolation from graphite, graphene, a one-atom-thick carbon sheet, has received significant attention due to its exotic electronic properties. However, since 2D graphene sheets have a gapless band structure, their practical applications in nanoelectronics are very limited. To overcome this limitation, many interesting studies have focused on graphene nanoribbons (GNRs), which can be produced by several methods including mechanical cutting of graphene sheets into finite widths [1–3] and graphene patterning [4, 5].

By their edge structures, GNRs are classified into zigzag GNRs (ZGNRs) and armchair GNRs (AGNRs). In a previous theoretical study [6, 7], ZGNRs were found to become half-metallic when an external transverse electric field was applied, while AGNRs are semiconductors, regardless of the length of the width. The electronic properties of GNRs generally change depending on their widths [7, 8], edge structures (such as edge terminations or defects) [9–23], and strain effects [22–27]. These properties can be modified to tailor their energy band gap and to expand the practical applications of graphene devices.

Recently, several basic reconstructions of the edges of GNRs have been studied [28, 29]. Koskinen *et al* [28] calculated the reconstruction of GNR edge structures using density functional theory calculations; they considered several geometries obtained through the reconstruction of zigzag and armchair edges and found some changes in their electronic structures. They also showed short segments of the predicted edges from transmission electron microscopy data on graphene [29]. However, most theoretical studies on edge-modified GNRs focus on ZGNRs with the adsorption or substitution of several atoms or chemical groups, reconstructions, and vacancies. To understand the properties of the edge structures of AGNRs, various edge-functionalization studies are required. Since the AGNRs with pentagonal-hexagonal edge structures (labeled (5,6)-AGNR) have the largest binding energy for the adsorption of hydrogen atoms due to the presence of dangling bonds [28], in this paper, we focus on (5,6)-AGNRs among the several possible edge reconstructed structures. By incorporating atom substitution, we have studied AGNR edge structures comprising five-membered heterocyclic compounds.

In the present study, we chose AGNRs with widths ($N$) of 20, 21, and 22 to represent three AGNR families, i.e. $3n-1$, $3n$, and $3n+1$ ($n$: positive integer), respectively. We investigated the structural and electronic properties of edge-functionalized (5,6)-AGNRs with substituted oxygen (O), nitrogen (N), sulfur (S), and selenium (Se) atoms. We then studied the electronic structures with consideration of all possible edge structures and analyzed the charge distribution of the subbands near the Fermi energy. As a result, we found that the electronic structure of the (5,6)-AGNRs depend on the number of electrons at the edge.

## 2. Computational details

We performed density functional theory calculations using the Vienna *ab initio* simulation package (VASP) [30, 31]. The cutoff kinetic energy was 400 eV and the ions were represented by projector-augmented wave potentials [32, 33]. A generalized gradient approximation (GGA) was employed to describe the exchange-correlation functional [34, 35]. The atomic positions of all structures were relaxed with residual forces smaller than 0.02 eV/Å. We used a 1 × 15 × 1 grid in the Monkhorst–Pack special *k*-points scheme centered at Γ ($k = 0$) for Brillouin-zone integration. To calculate the electronic band structures of the nanoribbons, 60 *k*-points were used in the direction of the AGNR axis. The distance between the GNRs was greater than 15 Å to avoid unwanted interactions between the adjacent GNRs. To find the suitable lattice constant of AGNR, we optimize the cell volume of the AGNRs. These lattice constants were found to converge the stress tensors to well within 0.5 kbar.

## 3. Results and discussion

First, we calculated the electronic structures of defect-free AGNRs to confirm the effect of edge-functionalized (5,6)-AGNRs: Band gaps of 0.12, 0.35, and 0.52 eV were obtained for $N = 20$, 21, and 22, respectively where *N* is the width index of AGNR. Following the previous paper of GNRs [7], the width index of AGNRs represents the number of dimer lines across the ribbon width. Left side plot of figure 1(a) shows defect-free AGNRs with width index $N = 21$. Small black arrows indicate all the dimer lines of AGNRs along the ribbon width. We considered the defective structures with one C–H bond missing at each of both edges of AGNR for three families of defect-free AGNRs with relaxed lattice constants. Those defective structures are usually denoted as the (5,6)-AGNRs; for example, (5,6)-21-AGNR with the width index $N = 21$ is shown in figure 1(b). The lattice constants decrease by ~2.5% and the electronic structures of (5,6)-AGNRs change significantly due to the effect on the edge structures. The relaxed lattice constants of the (5,6)-AGNRs are listed in table 1.

**Table 1.** Relaxed lattice constants, $d_{\text{lattice}}$, and band gaps, $E_g$, of defect-free AGNR, (5,6)-AGNR, and edge-functionalized (5,6)-AGNRs.

|  |  | Defect-free | (5,6) | NH-(5,6) | O-(5,6) | S-(5,6) | Se-(5,6) |
|---|---|---|---|---|---|---|---|
| 20-AGNR | $d_{\text{lattice}}$ (Å) | 4.280 | 4.174 | 4.171 | 4.163 | 4.212 | 4.232 |
| | $E_g$ (eV) | 0.12 | — | 0.47 | 0.69 | 0.63 | 0.62 |
| 21-AGNR | $d_{\text{lattice}}$ (Å) | 4.281 | 4.180 | 4.177 | 4.170 | 4.212 | 4.234 |
| | $E_g$ (eV) | 0.35 | — | 0.03 | 0.20 | 0.19 | 0.27 |
| 22-AGNR | $d_{\text{lattice}}$ (Å) | 4.280 | 4.187 | 4.183 | 4.177 | 4.219 | 4.235 |
| | $E_g$ (eV) | 0.52 | — | 0.45 | 0.32 | 0.26 | 0.18 |

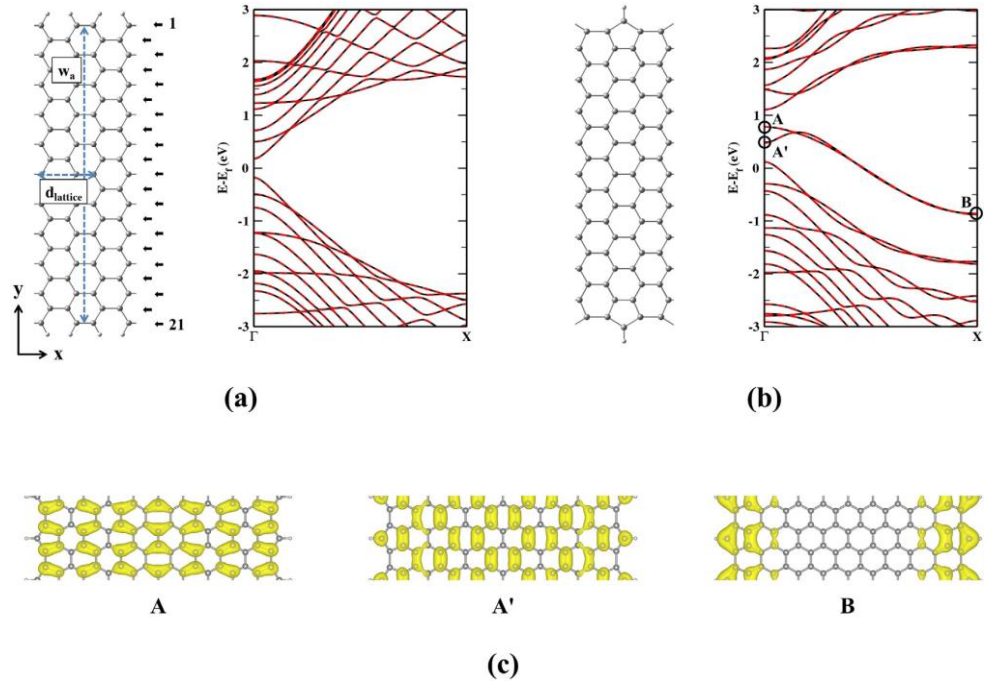

**Figure 1.** Optimized geometries and electronic structures of (a) 21-AGNR and (b) (5,6)-21-AGNR. The grey circles and small white circles represent carbon atoms and hydrogen atoms passivating the edge carbon atoms, respectively. The unit cell parameter and ribbon width are represented by $d_{\text{lattice}}$ and $w_a$, respectively. (c) Partial charge density plots of edge-related states denoted in (b).

Figure 1(c) shows the charge densities at $\Gamma$ and $X$ points of the subbands in the (5,6)-AGNRs. The electronic states marked as $A$ and $A'$ at $\Gamma$ show the extended states. In contrast, state $B$ at $X$ is the localized edge state in the forbidden energy region. To explain these states in the (5,6)-AGNRs, we introduce the concept of complex band structures. In Bloch's theorem, the eigenstates, $\psi_{n\mathbf{k}}(\mathbf{r})$, of the single-electron Schrödinger equation in a crystal satisfy $\psi_{n\mathbf{k}}(\mathbf{r}) = u_{n\mathbf{k}}(\mathbf{r})e^{i\mathbf{k}\cdot\mathbf{r}}$ where $u_{nk}(r)$ is a function that has the same periodicity as the crystal, *n* is the band index, and *k* is the wave vector. For an infinite crystal, the wave vectors should be restricted to real quantities by the Born–von Karman cyclic boundary conditions. However, the wave functions with complex wave vectors *k* are also solutions of Schrödinger equation in general. Near a crystal surface or interface, one can match a wave function with a complex wave vector $k = \boldsymbol{k}_r + i\boldsymbol{k}_i$ between the inside and outside of the crystal region, and thus decaying states occur at the surface. Here, $\boldsymbol{k}_r$ and $\boldsymbol{k}_i$ are the real and imaginary components, respectively. The complex band structure concept can be applied to graphene edges since the electronic properties of the edge states are associated with the band structure of infinite graphene. This concept enables us to predict the decay behavior of the edge state in semi-infinite graphene, combined with the complex band structure of bulk graphene. We can estimate the decay length using, $\lambda = 1/|k| = |E_d - E_{c(v)}|^{-\alpha}$ where $\alpha > 0$. Here, $E_d$, $E_v$ and $E_c$ are the energy levels of the defect state, the valence and conduction band edges, respectively. The complex band concept for graphene can be applied to understand the localized states at the graphene edges [36]. Briefly speaking, the decay length (from the edge to bulk) is short when $E_d - E_{c(v)}$ is large. Because the forbidden energy region is largest at the *X* point for the GNR, the decay length at *X* is shortest and so the localization character at the ribbon edge is strongest at *X*. The general properties of complex band structures can be derived from the bulk band structure of graphene. The decay patterns of the (5,6)-AGNRs are associated with the complex band structure of bulk graphene. Our results concerning the electronic structures are similar to those reported previously [36].

To investigate the effect of the edge structures of reconstructed AGNRs, we considered structures with substitution of a CH unit at the edge with other atoms; the substituted atoms were selected to maintain the pentagonal structure at the edges of AGNRs. Pyrrole, furan, thiophene, and selenophene are well-known heterocyclic molecules with pentagonal structures. Therefore, we investigated edge-functionalized (5,6)-AGNRs including O, S and Se atoms and the NH group. First, we calculated the optimized structures and total energies of the isolated molecules in vacuum; the optimized structures are shown in figure 2.

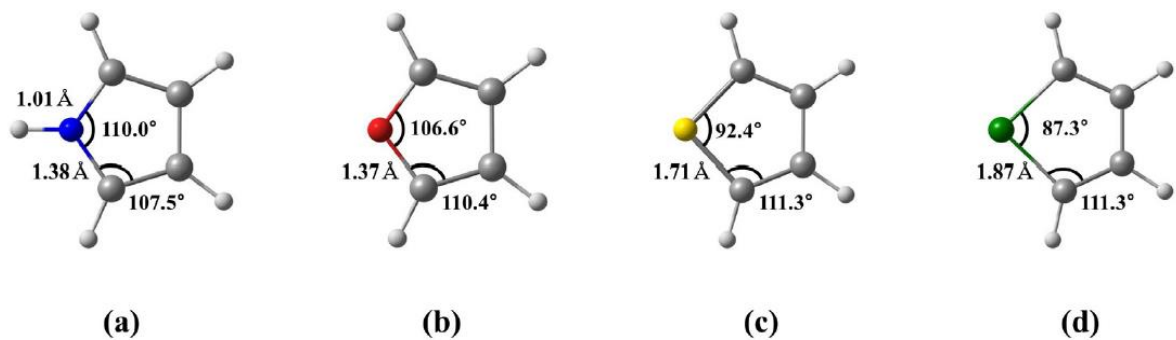

**Figure 2.** Structures of the five-membered rings. (a) Pyrrole, (b) furan, (c) thiophene, and (d) selenophene. The grey, white, blue, red, yellow, and green balls are the carbon, hydrogen, nitrogen, oxygen, sulphur, and selenium atoms, respectively.

Then, we considered edge-functionalized (5,6)-AGNRs and optimized these structures using relaxed lattice constants. After optimization, the O- and Se-substituted AGNRs had the shortest and longest lattice constants, respectively. Although we have performed calculations for all three families of AGNRs, we explain the details of the analysis of the (5,6)-21-AGNR cases, belonging to the (5,6)-3n-AGNR family, only in this paper. Other structures such as (5,6)-20-AGNR and (5,6)-22-AGNR have similar trends to (5,6)-21-AGNR.

In the optimized structures of the (5,6)-21-AGNRs, both the distance between the impurity (N, O, S, or Se) and an adjacent carbon atom and the C–I–C angle (I: the impurity atom) increase when incorporated into the AGNR edge, while the I–C–C angle decreases by about 8° (see figure 3), as compared to the corresponding gas-phase molecule (see figure 2). To study the effects of edge-functionalization of (5,6)-AGNRs, we calculated electronic structures for all edge-functionalized (5,6)-AGNRs; Middle panels and bottom panels in figure 3 show the band structures and the charge densities of the impurity bands (blue lines in the band structures) at the $X$ point corresponding to localized edge states of (5,6)-21-AGNRs containing NH-, O-, S-, and Se-substituted edges, respectively.

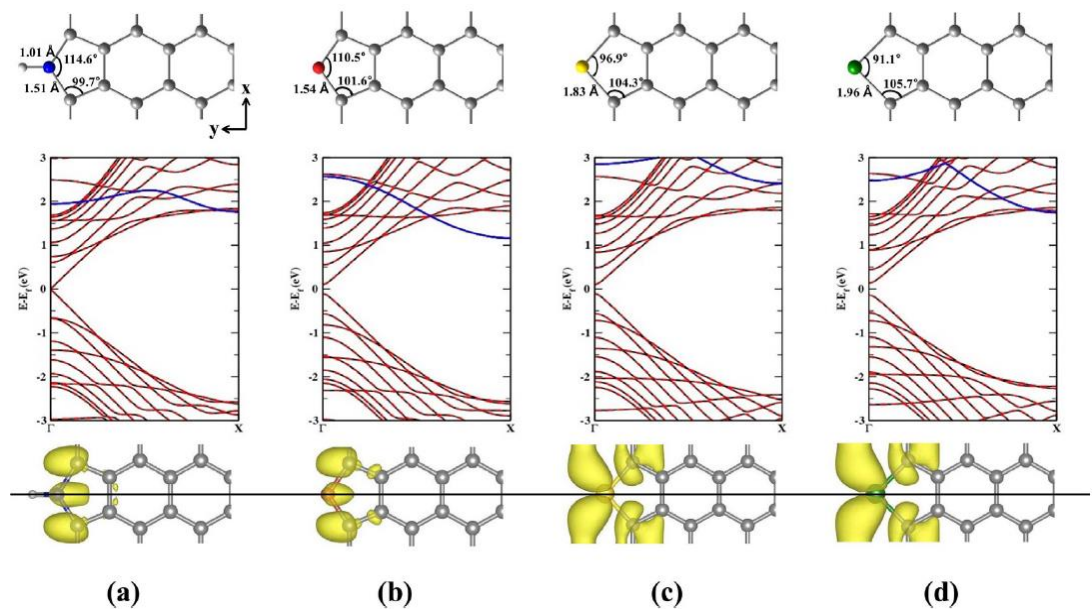

**Figure 3.** Optimized geometrical structures, band structures, and partial charge density plots of impurity bands at the $X$ point of edge-functionalized (5,6)-21-AGNR with (a) the NH group (blue), (b) the O atom (red), (c) the S atom (yellow), and (d) the Se atom (green) substitution. Blue line in the band structures represent the impurity bands from functionalizing atoms of edge-functionalized (5,6)-AGNRs.

We found two changes in the main properties of the electronic structures upon substitution. One is a change in the band structure at the Γ point. That is, the band structures of edge-functionalized (5,6)-21-AGNRs at the Γ point become similar to those of the defect-free 20-AGNR as shown in figure 4(a); this tendency is more common in the conduction band than in the valence band. To find the changes of the edge-functionalized (5,6)-AGNRs, we investigated subbands in the energy range of −1.0 to 1.0 eV with respect to the Fermi level. In the case of edge-functionalized (5,6)-21-AGNRs, the crossing between the second and the third subbands which are located upward (or downward) from the Fermi level occur when going from Γ to X, (see figure 4(a)). This behavior differs from the case of defect-free 21-AGNR, where such crossing occurs between the first and the second subbands, as shown in figure 4(b).

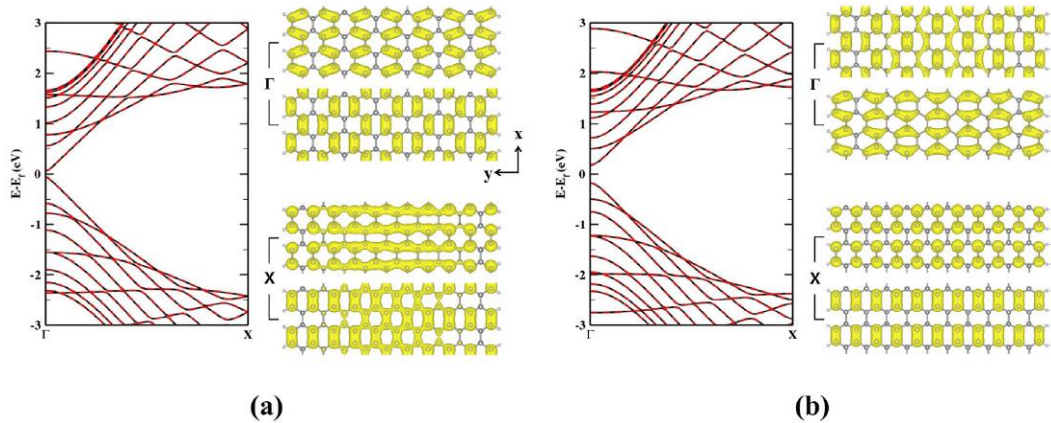

**Figure 4.** Band structures and partial charge plots at Γ and X of the subbands near the Fermi level of both defect-free (a) 20-AGNR and (b) 21-AGNR. Upper panels of partial charge plots for each of Γ and X show the lowest conduction bands and lower panels represent the highest valence bands.

The other is a difference in the band gap sizes of (5,6)-AGNRs. The band structures of the (5,6)-AGNRs have metallic properties (see figure 1(b)), whereas substitution of the impurity in (5,6)-AGNRs restores band gaps similar to those of defect-free AGNRs. All (5,6)-AGNRs show direct band gaps at Γ. The band gap size is affected by the width of the AGNR and the type of the substituted atom at the edge of the AGNR. It is clear that the energy gaps of the reformed AGNRs can be tuned by the width variation and edge functionalization. In defect-free AGNRs, the band gap size is in the order of $(3p + 1)$-AGNR > $(3p)$-AGNR > $(3p + 2)$-AGNR ($p$: positive integer). In contrast, the band gaps of the edge-functionalized (5,6)-AGNR at the edge vary in the order of (5,6)-20-AGNR > (5,6)-22-AGNR > (5,6)-21-AGNR. The changes in the energy gaps support that the band structures of edge-functionalized (5,6)-$N$-AGNRs resemble that of defect-free $(N-1)$-AGNR. The band gaps of (5,6)-AGNRs substituted with NH, O, S, and Se are listed in table 1. Note that the case of edge-functionalized (5,6)-AGNRs by Se atoms is an exception since those AGNRs do not follow the aforementioned trend for band gap size, showing that (5,6)-20-AGNR > (5,6)-21-AGNR > (5,6)-22-AGNR.

A possible reason for such dramatic changes in the band structures of substituted (5,6)-AGNRs is that the number of electrons at the edge decreases through reconstruction of the AGNRs with pentagonal edge structures, which breaks the bonding symmetry of the resonance in the AGNRs. By substituting a CH unit with NH, O, S, or Se in (5,6)-$N$-AGNRs, the number of electrons at the edge increases to be equal to that of the defect-free ($N$-1)-AGNR, which can be explained below.

We consider only the $N$ = 21 case without loss of generality. The 21-AGNR configuration for a narrow unit with 2CH at each edge can be decomposed into the following: (21-AGNR with H) = 2CH + (19-AGNR w/o H) + 2CH = 2CH + 2C + (18-AGNR w/o H) + 2CH. If 2CH at both edges are replaced by, say, O atoms, the configuration changes into O + 2C + (18-AGNR w/o H) + O. In terms of the number of electrons at the edge, it can be understood as (6$e$) + (2 × 4$e$) + (18-AGNR w/o H) + (6$e$), which, after rearranged, becomes (10$e$) + (4$e$) + (18-AGNR w/o H) + (−4$e$) + (10$e$) = (10$e$) + (18-AGNR w/o H) + (10$e$): The latter electronic configuration gets back to 2CH + (18-AGNR w/o H) + 2CH = 20-AGNR with H. Thus, we can conclude that in terms of the number of electrons at the edge, the electronic configurations of functionalized (5,6)-$N$-AGNR with substitution atoms resemble that of defect-free ($N$-1)-AGNRs at Γ. Essentially, the number of electrons at the edge determines the band structures of AGNRs. This means that the orbital bonding between the carbon atoms and substituted atoms at the edge of (5,6)-$N$-AGNRs is similar to the bonding between carbon atoms in defect-free ($N$-1)-AGNRs. However, the band structures of edge-functionalized (5,6)-$N$-AGNRs at $X$ show the original ones of the defect-free $N$-AGNR, as manifested from band structures of figure [3](#) and that of figure [4(b)](#) at $X$. As mentioned before, the coupling between the defect state and extended state is enhanced as $E_d − E_{v(c)}$ is small. At the $X$ point, the coupling is weak and the decay length of the edge state is short in the GNR. We conclude that the electronic structures at $X$ are not influenced by the edge structure because the defect states at the edge are practically decoupled to the extended GNR states at $X$.

We also found an impurity band in all the band structures that is caused by the incorporation of NH, O, S, and Se atoms (blue line in figure [3](#)); the band descended toward the Fermi level from Γ to $X$. The edge state of the impurity band at Γ is located in the conduction band structure and the charge density at $X$ is located at the GNR edge (see the bottom panel of figure [3](#)). The partial charge density plots corresponding to the localized edge state are symmetric with respect to the plane perpendicular to the GNR axis direction.

To more accurately analyze the band structures, we plotted the charge distribution of subbands near the Fermi level. As an example, the charge distributions of the (5,6)-21-AGNRs are presented in figure 5. Figures 5(a) and (b) show the partial charge distribution corresponding to the two subbands of (5,6)-21-AGNRs with different functionalized edges at Γ and X, respectively. In figure 5, the upper panels for each of Γ and X points represent the charge distributions of the lowest conduction bands, while the corresponding lower panels give those of the highest valence bands. The subbands at Γ (see figure 5(a)) show two states: one represents the vertical bonds between carbon atoms, while the other represents the bonds parallel to the periodic direction.

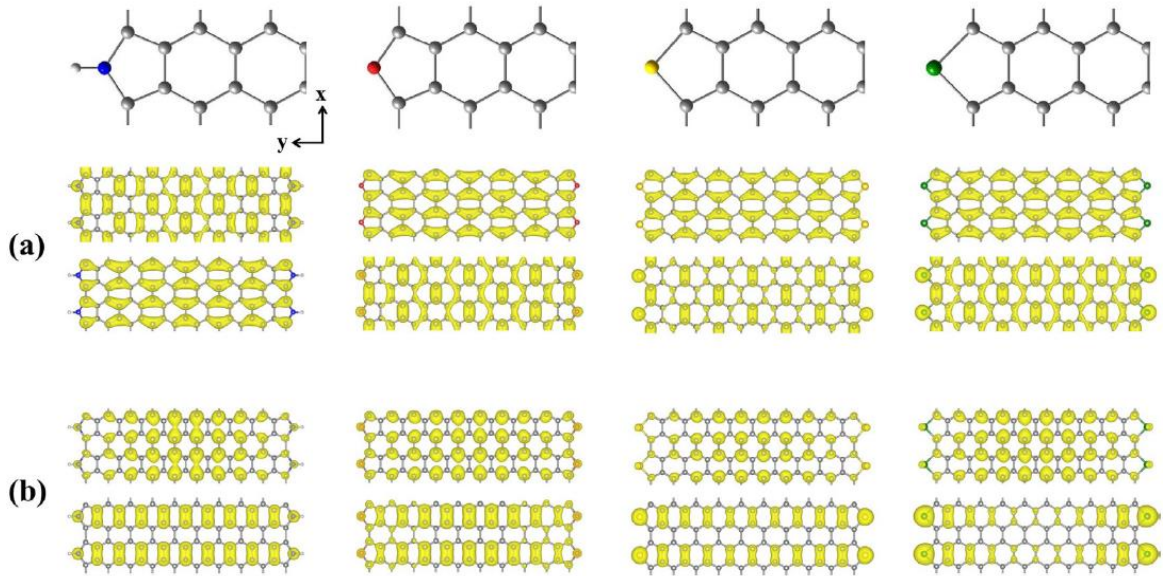

**Figure 5.** Partial charge plots of (a) Γ and (b) X of the subbands near the Fermi level in the band structures of NH-, O-, S-, and Se-functionalized (5,6)-21-AGNR. Upper panels and lower panels for each of Γ and X points represent partial charge plots of the lowest conduction bands and the highest valence bands, respectively.

As expected due to their band structures, the pattern of the charge distributions of the subbands at the Γ point of edge-functionalized (5,6)-21-AGNRs containing the O, S, Se atom or the NH group are similar to that of defect-free 20-AGNR. It is found that the subbands in the energy range of about −1.0 to 1.0 eV with respect to the Fermi level show only the $p_z$ orbital characters of the carbon atoms and substitutional atoms of the AGNRs. At the X point, the electronic density plots of (5,6)-AGNRs show the bonds parallel to the periodic direction. Similarly to the Γ point, only $p_z$ orbitals of all the atoms contribute to the subbands near the Fermi level at the X point. Partial charge plots for subbands at X of edge-functionalized (5,6)-21-AGNRs in figure 5(b) resemble those at X of defect-free 21-AGNR in figure 4(b) from the viewpoint of bonding or antibonding character between carbon atoms of AGNRs. Note that the symmetry of geometrical edge configurations of defect-free N-AGNRs are the same as that of edge-functionalized (5,6)-N-AGNRs; that is, the edge configurations of both defect-free 21-AGNRs and edge-functionalized (5,6)-21-AGNRs are symmetric. These symmetric edge configurations may partly explain why at the *X* point the edge-functionalized (5,6)-21-AGNR shows the same pattern as defect-free 21-AGNR for the band structures and partial charge densities.

## 4. Conclusions

In conclusion, we examined the structural and electronic properties of the edge-functionalized pentagonal-hexagonal configurations of AGNRs, i.e. (5,6)-AGNRs, substituted with several different atoms using first-principles calculations. Motivated by pyrrole, furan, thiophene, and selenophene molecules, O, S, and Se atoms and the NH group were selected to functionalize the pentagonal edge of the AGNRs. After obtaining the optimized structures, the differences in the electronic properties of the edge states were analyzed; they were dependent on the width size of the AGNRs. By chemical functionalization, the number of electrons at the edge can be tuned, which affects the electronic structures. At the $\Gamma$ point, the electronic properties of edge-functionalized (5,6)-*N*-AGNRs are similar to those of defect-free (*N*-1)-AGNR, whereas they are similar to those of defect-free *N*-AGNR at the *X* point. Thus, edge functionalizing atoms can modify the electronic structures of the metallic edge-reconstructed (5,6)-AGNRs with substitution of CH.

## Acknowledgments

This research was supported by Nano•Material Technology Development Program (2012M3A7B4049888) through the National Research Foundation of Korea (NRF) funded by the Ministry of Science, ICT and Future Planning (MSIP), and Priority Research Centers Program (2010-0020207) through NRF funded by the Ministry of Education (MOE). G.K. thanks the financial support of the Basic Science Research Program through MSIP/NRF (2010-0007805).